\documentclass[journal=jacsat,manuscript=article]{achemso}
\usepackage[version=4]{mhchem}
\usepackage{nicefrac}
\usepackage{siunitx}
\usepackage{graphicx}
\usepackage{lineno}
\author{Emanuele Grifoni}
\affiliation[ETH]
{Department of Chemistry and Applied Biosciences, ETH Zurich, c/o USI Campus, Via Giuseppe Buffi 13, CH-6900 Lugano, Ticino, Switzerland}
\alsoaffiliation[USI]
{Institute of Computational Science, Università della Svizzera italiana (USI), Via Giuseppe Buffi 13, CH-6900, Lugano, Ticino, Switzerland}
\author{GiovanniMaria Piccini}
\affiliation[ETH]
{Department of Chemistry and Applied Biosciences, ETH Zurich, c/o USI Campus, Via Giuseppe Buffi 13, CH-6900 Lugano, Ticino, Switzerland}
\alsoaffiliation[USI]
{Institute of Computational Science, Università della Svizzera italiana (USI), Via Giuseppe Buffi 13, CH-6900, Lugano, Ticino, Switzerland}
\author{Michele Parrinello}
\email{parrinello@phys.chem.ethz.ch}
\affiliation[ETH]
{Department of Chemistry and Applied Biosciences, ETH Zurich, c/o USI Campus, Via Giuseppe Buffi 13, CH-6900 Lugano, Ticino, Switzerland}
\alsoaffiliation[USI]
{Institute of Computational Science, Università della Svizzera italiana (USI), Via Giuseppe Buffi 13, CH-6900, Lugano, Ticino, Switzerland}
\alsoaffiliation[IIT]
{Italian Institute of Technology, Via Morego 30, 16163 Genova, Italy}

\title[]{Tautomeric equilibrium in condensed phases}

\begin{document}

\begin{abstract}
We present an ab initio molecular dynamics (MD) investigation of the tautomeric equilibrium for aqueous solutions of glycine and acetone at realistic experimental conditions. Metadynamics is used to accelerate proton migration among tautomeric centers. Due to the formation of complex water-ion structures involved the proton dynamics in the aqueous environment, standard enhanced sampling approaches may face severe limitations in providing a general description of the phenomenon. Recently, we developed a set of Collective Variables (CVs) designed to study protons transfer reactions in complex condensed systems [Grifoni et al. PNAS, 2019, 116(10), 4054-4057]. In this work we applied this approach to study proton dissociation dynamics leading to tautomeric interconversion of biologically and chemically relevant prototypical systems, namely glycine and acetone in water. Although relatively simple from a chemical point of view, the results show that even for these small systems complex reaction pathways and non-trivial conversion dynamics are observed. The generality of our method allows obtaining these results without providing any prior information on the dissociation dynamics but only the atomic species that can exchange protons in the process. Our results agree with literature estimates and demonstrate the general applicability of this method in the study of tautomeric reactions.
\end{abstract}

\section{Introduction}
Tautomerism is a chemical phenomenom of great interest in which two isomeric molecular structures can interconvert\cite{Izvekov2005AbRevisited,Zhang2012,Pomes1998FreeMolecules,Rodriquez2001ProtonTriglycine,Hassanali2013ProtonGossamer,Jong2018AWater}. Although any reaction involving changes in the isomers connectivity is in principle tautomeric, only those reactions that imply an intramolecular relocation of protons are called tautomeric. 

The study of how inter and intramolecular proton transfers proceed and a quantitative assessment of the equilibrium constants between different tautomers are still missing despite the fact that these reactions are at the heart of several biophysical processes such as protein folding or enzymatic reactions. 
%This is mostly due to the fact that solvent is an important actor in these processes since this transfer occurs through the formation and diffusion of ionized water molecules and standard quantum chemical methods are met with significant difficulties. Water with its complex and fluctuating hydrogen bond network is difficult to describe with standard approaches especially if it participates actively to the reaction.
This is mostly due to the fact that water is an important actor in these processes and standard quantum chemical methods are met with significant difficulties. Protons are transferred between sites through the formation and diffusion of ionized solvent molecules and water with its complex and fluctuating hydrogen bond network is difficult to describe with standard approaches especially if it participates actively to the reaction.
Ab-initio molecular dynamics (MD) simulations are in principle better suited at describing these processes since the complex water dynamics is explicitly included. Applications of ab-initio MD to this process are however hampered by the fact that tautomeric transitions are rare events on the simulation timescale. This requires the use of enhanced sampling methods able to accelerate configurational space exploration.\cite{Bernardi2015}.

Among these methods, one very popular class is based on the identification of the slow degrees of freedom of the reaction\cite{Torrie1977,Laio2002,Valsson2014}. These degrees of freedom, or Collective Variables (CVs), are functions of the atomic coordinates and must be properly chosen. Sampling is then accelerated by adding to the physical energy landscape an external bias potential that is function of the chosen CVs. The scope of the bias is to enhance the CV fluctuations and to encourage the system to explore new states. 

However, finding good CVs can be challenging and in particular for reactions that involve proton transfer in water of where one has to deal with intermediate species in which a hydronium or hydroxyl ion is present in the solvent. The process by which the excess or defect of proton migrates in the solvent is usually refered to as the Grotthuss mechanism\cite{Agmon1995}. In this mechanism a charge defect migrates rapidly through the water network without a major rearrangement of the atomic positions. Several structures are associated to these charged species and the identity of the participating molecules change continuously\cite{Uberschub-protonen1968,Zustand1907,Marx1999,Hulthe1997,Iyengar2005}. Thus the description of this phenomenon in terms of a simple function of the atomic coordinates is not straightforward.
%, it is not possible to associate specific identities or geometries to the molecules involved. Moreover, due to their elusive and high fluctional nature, a large ensemble of different structures are associated to these charged species\cite{Uberschub-protonen1968,Zustand1907,Marx1999,Hulthe1997,Iyengar2005}. It is indeed known that a varying number of water molecules must be considered as active participants to the structure of the charge carriers\cite{Giberti2014,Kreuer2000,Marx1999}. Because of this structural behaviour a description of these species in terms of atomic coordinates is not simple. 
In a recent work we have introduced a new set of CVs\cite{Grifoni2019MicroscopicEquilibrium} that solves this problem and accelerates the study of reactions involving proton transfer events. In this paper we make use of this methodological advancement and study the tautomerism of glycine and acetone in water, two systems that are representative of a wider class of more complex systems. 

\begin{figure}
    \centering
    \includegraphics[width=1.\textwidth]{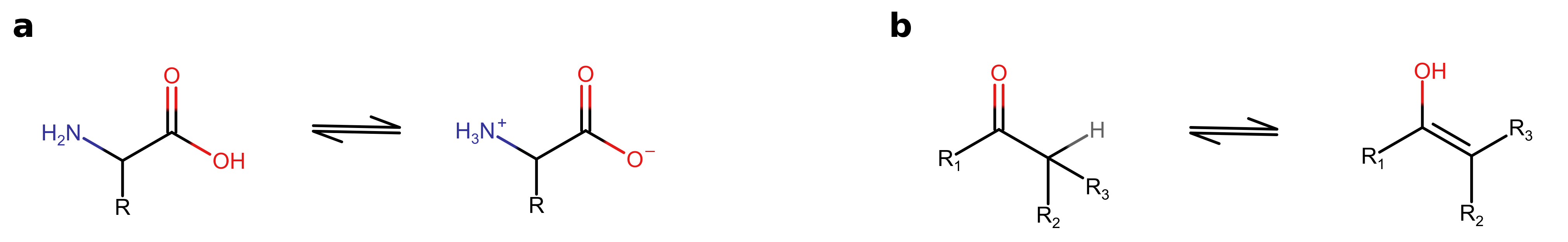}
    \caption{Zwitterionic equilibrium of a generic amino-acid (a) and keto-enol equilibrium of a generic ketone (b).}
    \label{fig:zwitterionic}
\end{figure}

Glycine is a small amino acid, the chemical building blocks of proteins, thus its investigation is a first step towards a systematic study of acid-base equilibrium in proteins (see Fig.~\ref{fig:zwitterionic}a). Enzimatic activity, conformational equilibrium and many other properties are strongly related to the $pKa$ of their monomeric units and must be, therefore, investigate carefully. 

%In proteins, due to the acid-base properties of many residues, the zwitterionic equilibrium (see Fig.~\ref{fig:zwitterionic}a) leads to a great variety of tautomeric combinations whose understandig is of great biochemical relevance. 

%It is not surprising that their properties are altered by the presence of local charges along their chains and their chemical behaviour affected by the protonation state of their monomers\cite{Zacharis2002ControlBuffers,Xu1996PHSolvents}. Thus the understanding of the mechanisms involved in these macromolecules has been the subject of several studies.

Analogously, acetone is the simplest molecule among ketones, a class of compounds of great importance in chemistry. These molecules can tautomerize in their enol forms but, while in amino acids there is an equilibrium between its canonical and zwitterionic form, here the charge imbalance due to the proton transfer is not a stable state and therefore it is instantaneously compensated by a reorganization of the electronic structure (see Fig.~\ref{fig:zwitterionic}b). 
%Then, this reaction has been chosen to represent non-zwitterionic tautomerizations. Due to the weak acid-base properties of the groups involved, here the charge imbalance caused by the proton transfer is stable. However, other tautomers pairs undergo a reorganization of their electronic structures during the process in order to compensate this charge separation.
In keto-enol equilibria and in other non-zwitterionic tautomerizations a hydrogen atom and a double bond migrate simultaneously.
%As a representative case of this class of tautomeric processes we studied the interconversion of acetone from its keto form to the isopropenl enol variant. Keto-enol tautomerism regulates many processes in chemistry and biochemistry. 
This class of tautomeric reactions underlies not only several biochemical processes like DNA base mutations and aldose-ketose interconversion of sugars, but they are also crucial in many others like supramolecular assembly, thermo and photochromism, and also in polymeric growth\cite{Breton2009Iminothiol/thioureaAssembly,Chapelet-Letourneux1961ACoordinates,Suarez1998SupramolecularMesophase,Cederstav1994InvestigationsAcetaldehyde,Yoshii2014ConjugatedBoron-complexation}. 

%In the latter field one can quote mutations of DNA bases, aldose-ketose interconversion of sugars and are also used to store energy taking advantage by the difference in stability between the keto and the enol forms.

%Identifying these thermodynamic states and assigning to them a statistical weight proportional to their stability is an useful tool not only for extracting acid-base equilibirum, but even for studying how they affect all their other properties.  
\section{Methods}
As previously mentioned, these reactions imply intermediate states involving solvated water ions and the description of these species as a function of their atomic coordinates is difficult. The varying number of water molecules that must be considered participating in the structure of these charge carriers\cite{Giberti2014,Kreuer2000,Marx1999} makes impossible to relate these species to definite chemical structures. In a previous work\cite{Grifoni2019MicroscopicEquilibrium} we developed a new set of CVs that circumvents these problems in two steps: the excess and defect of protons are looked at as charge defects and all the solvent molecules are considered like a single macromolecule that collectively hosts one or more charge defects. This point of view allows identifying these species as charge anomalies without any reference to a particular structure. 
%Having taken these steps we circumvent the problem of associating fixed structures to the protonated species. Reactants and products are in this way defined according to local anomalies in the number of protons irrespective of their specific geometries.

The reference charge for water is that of a neutral molecule and thus we look for deviations from this value. Then, since we consider the solvent as a single reactant, its overall charge is given by the sum of all its molecular contributions. Similarly we identify for each solute molecule the moieties that are able to release or accept protons and again we look for anomalies relative to their reference states. In this procedure any reference to specific geometries is lost and reactants and products are defined according to local anomalies in the number of protons.

%Because of the fluctional nature of these charge carriers, their structures can be found in a virtually infinite number of configurations. Fixing the number of water molecules participating to these structures is not possible and therefore their shape and size are not good descriptors. Because of this, standard approaches fail to provide general descriptors able to identify these metastable species. However, the excess or defect of protons can be looked at as charge defects making possible to identify these species as charge anomalies without any reference to a particular structure. In a previous work, starting from this considerations, we have developed two CVs\cite{Grifoni2019MicroscopicEquilibrium} for the study of reactions involving relocation of protons. We define for each water molecule a reference state that in this case is the neutral one and we assigned the charge due to the proton excess or defect to the nearest water molecule. Then we consider the solvent as a single reactant and we look at  each of the overall charge. Similarly we identify for each solute molecule the moieties able to release or accept protons and again we look for anomalies relative to their reference states.

In order to define these charge anomalies in a precise way, we tessellate the entire space with Voronoi polyhedra centered on these sites and monitor the total charge that each polyhedra contain. See SI Appendix for details. If the deviation $\delta_i$ of the charge evaluated from the reference value is non zero, this defines a charge defect. 
%%%% tautomers are distinguished by the number of hydrogen atoms that contain. 

%The second one ensures a very precise and unambiguous way in the evaluation of this density.

%%%%%%%%%%%%%%%%%%%%%%%%%%%%%%%%%%
Then the CVs are defined as
\begin{equation}
    s_p=\sum_{k=0}^{N-1}2^k \cdot \sum_{i \in k} \delta_i,
    \label{eq:sp}
\end{equation}

\begin{equation}
    s_d=-\frac{1}{2}\sum_{i,m>i}r_{im}\cdot \delta_i \cdot \delta_m,
    \label{eq:sd}
\end{equation}
where $k$ is an index running over the inequivalent species, $i$ and $m$ run all over the site indexes and $r_{im}$ is their distance. See SI Appendix for details.
%This last transformation ensures that any tautomeric combination occupies a different position in the CV space. See SI Appendix for details.
The first CV, $s_p$, returns a value that uniquely identifies every tautomeric combination, the second one, $s_d$, measures the the distance between the sites that have exchanged a proton and therefore that are not in their reference states.

The systems studied were composed by a molecule of glycine and another one of acetone both solvated by 32 water molecules. The thermodynamic states chosen as starting points and references are the zwitterionic and the keto form respectively. Then $s_p$ has been used to enhance the transfer of a proton between two different sites while $s_d$ to enhance charge separation and accelerate the diffusion of charge carriers inside the solvent.

Ab-initio MD simulations have been used in combination with well-tempered metadynamics\cite{Laio2002,Barducci2008}. We used the CP2K package\cite{Vandevondele2005} patched with PLUMED2 \cite{Tribello2014PLUMEDBird}, an open-source plugin for enhanced-sampling. Exhaustive computational details can be found in the SI Appendix.
\section{Results}
\begin{figure}[htb]
    \centering
    \includegraphics[width=1.\textwidth]{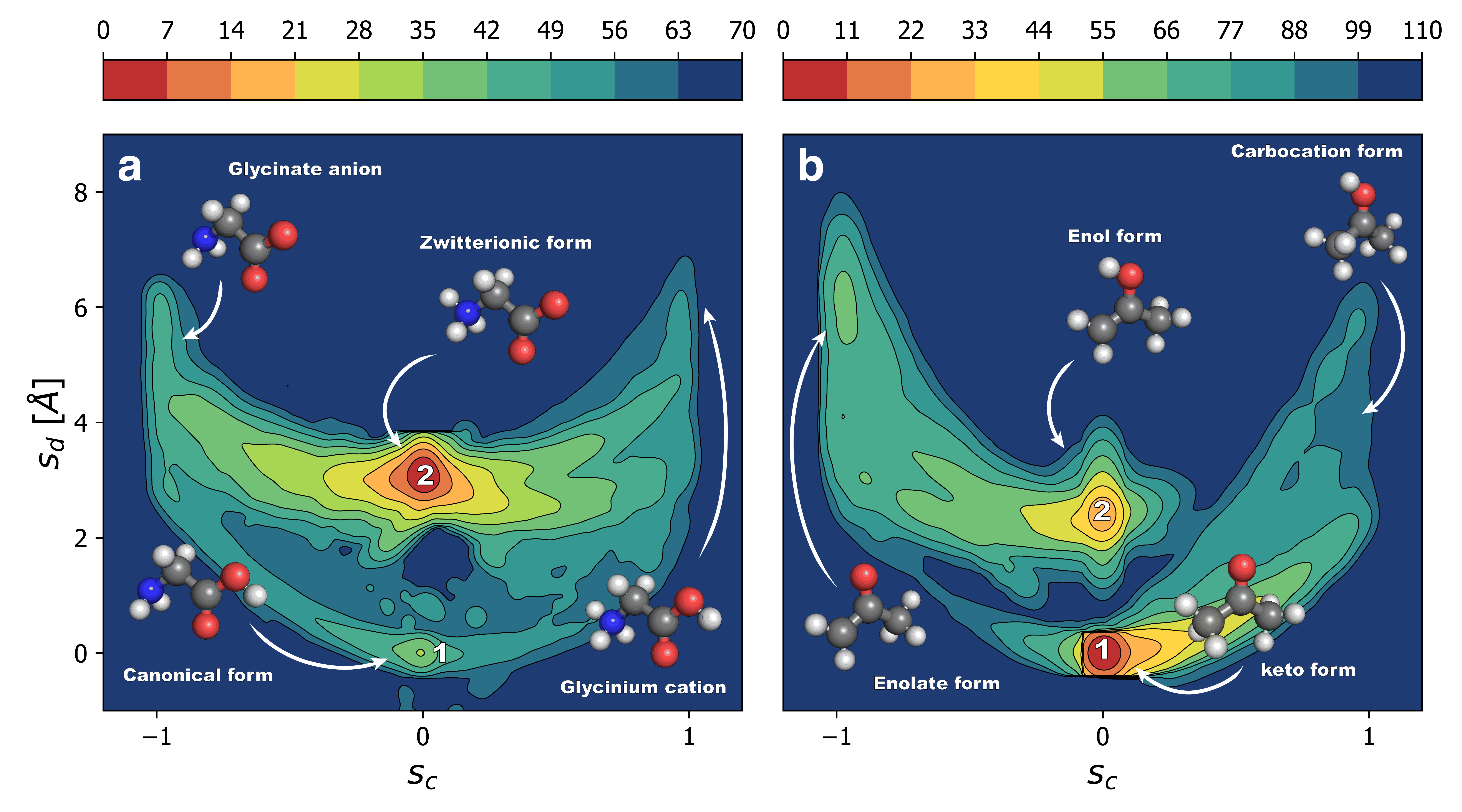}
    \caption{Free energy surfaces along $s_c$ and $s_d$ of glycine (panel a) and acetone (panel b) in aqueous solution. Along $s_c$ we can discern the two ionic and the two neutral states. The two tautomeric structures have both a value of $s_c$ equal to 0 while at -1 and +1 are located the two transient species, glycinate and glicinium ions respectively. Along $s_d$ is reported the distance between the sites that have exchanged a proton. Color bars indicate the free energy expressed in \si{kJ \cdot mol^{-1}} units (please note the different z-scales for the two plots).}
    \label{fig:fes}
\end{figure}

We have performed metadynamics calculations using the variables described in the Eq.~\ref{eq:sp} and \ref{eq:sd}. However, in order to extract more chemically meaningful, we found more expressive if we perform a reweighting\cite{Tiwary2015AMetadynamics} and project the outcome on two new variables. One is the total charge on the solute molecule $s_c$ and the other is $s_d$ as in Eq.~\ref{eq:sd} where we replaced, in the glycine case, the reference state with the canonical form. The Free Energy Surfaces (FESs) as a function of $s_c$ and $s_d$ are shown in Fig.~\ref{fig:fes} while those along $s_p$ and $s_d$ can be found in the SI appendix.

\subsection{Glycine}
%The metadynamics simulation has been biased along the CVs defined in the Method section, however a description of the system in terms of the total charge  of the solute molecule ($C$) and the charge distance ($s_d$) gives a more physically transparent description. In the same spirit, here we have used the canonical form as our reference state. The Free Energy Surface (FES) reweighted\cite{Tiwary2015AMetadynamics} along $C$ and $s_d$ is shown in Fig.~\ref{fig:fes}a while the FES as a function of $s_p$ and $s_d$ can be found in the SI appendix.

%In this process reactants and products are both neutral while their interconversion must occur through two ionic structures. Then, in this representation the two tautomeric structures have both a value of $C$ equal to 0 while at -1 and +1 are located the two transient species, glycinate and glicinium ions respectively.

The FES in Fig.~\ref{fig:fes}a exhibits two local minima located at $s_c=0$. The lower one, basin 1, is very shallow and high in energy and corresponds to the canonical form while the the other, basin 2, is much deeper and refers to the zwitterionic form. Fluctuations from the canonical basin toward negative values of $s_c$ means that the glycine is releasing a proton from its carboxylic group to a nearby water molecule while positive $s_c$ values mean that its amino group is taking a proton from the solvent. The formation and diffusion of these water ion structures is reflected in an increase of the $s_d$ value. Same considerations can be made in the case of the zwitterionic tautomer described by the basins 2. In this case the anionic structure is reached by losing a proton from the amino group and the cationic one by protonation of the carboxyl oxygen atoms. 

%This implies that in proximity of the state 1 and 2 the evolution of $s_d$ is strictly correlated with the variable $C$ and this explains the sickle-shaped basins. 
%However different, NH and OH bonds are close in energy and

%Same considerations can be made for the canonical form described by the basins 2. Here the basin is less pronounced and therefore small differences between fluctuations toward the two FES sides are more evident. We can see clearly the preference of the canonical form to have fluctuations toward the left side of the FES than the right one. This implies that the covalent bond between the hydrogen  and the carboxyl oxygen atoms are slightly weaker than the bond between a proton and the nitrogen atom.

In addition, this FES provide also useful insight about the interconversion mechanism. We can distinguish the two reactive pathways involving the formation of two ionic structures, glycinate and glicinium ions. In these transient species only one of the two glycine groups has reacted making them  positively or negatively charged. It is not surprising that, in this representation, the two main basins are not connected by a straight reactive pathway. In fact, the interconversion can only be reached passing through other higher energy intermediates and their water counterions. This reflects the lack of a direct hydrogen bond pathway between the ammino and the carboxyl group that does not pass through the solvent. The distance between the glycine two ends is too large to allow a direct proton jump and the reaction proceeds via extramolecular low energy paths through the solvent hydrogen bond network. 
%Due to the finite size effect, we could not do an accurate analysis of the hydrogen bonds network connecting the glycine two ends. However proton wires are still visible and it is also observable that the charge transfer takes place through them. 
We can also observe that both glycinium cation and glycinate anion are rather unlikely and once one of them is fully formed our system relax quickly in one of the two isomeric forms. Furthermore we can see that, even though small differences can be still observed, the two reactive pathways are roughly equivalent in terms of energy. This approximate left-right symmetry reflects the comparable strength of the bond that we need to break or form in order to move toward one of the two ionic structures. This means that, despite a lower barrier in proximity of the anionic structure, the tautomerization can occur either through the anionic or the cationic intermediate states.

Finally, we found that the zwitterionic structure is lower in energy than the canonical form by 35 \si{kJ \cdot mol^{-1}}, value in agreement with the literature estimates\cite{DawsonR.M.C.ElliottDaphneC.ElliottW.H.Jones1960}.

%Moreover, for the basin 3, we observe a double minimum along $s_d$. This is caused by the the presence of two inequivalent carboxylate group oxygen atoms able to bond the proton\cite{Lee2010BarrierlessGlycine-H2O9}.  

\subsection{Acetone}
Similar considerations can be made for the acetone FES reported in Fig.~\ref{fig:fes}b. The states 1 and 2 identify the keto and enol form respectively while at the two sides we find their transient structures. 

Contrary to the previous example, here we can immediately notice the absence of a left-right symmetry. From the keto form (basin 1) the amplitude of the fluctuations are larger toward the right side compared to those pointing to the opposite direction. These fluctuations are due to the protolysis of the carbonyl group and methyl group respectively. Similarly, the enol form (basin 2) shows the same behaviour and the proton transfer to the carbonyl group is more favoured than that to the methylidene group. This is explained by the different strength of the bonds involved in this tautomeric process. The non-polar nature of carbon-hydrogen bonds makes moving a proton between a methyl group and a nearby water molecule harder than with the oxygen-hydrogen bonds and this leads to inequivalent thermodynamic paths and not symmetric fluctuations around their minima.

In agreement with the literature\cite{Heinrich1986SubstituentTautomers,Chiang1989TemperaturePhase}, the basin corresponding to the keto form is much deeper than the enol form with a difference in free energy equal to 34 \si{kJ \cdot mol^{-1}}. Furthermore, we can see that these transient structures are higher in energy then those in glycine. Compared to glycine, here the presence of charges leads to thermodynamic states even more rarely visited and the solute molecule will spend most of its time in states corresponding to neutral structures.
\section{Conclusions}
The results presented in this work demonstrate the general applicability of this method in the study of tautomeric interconversions. The glycine case represents those systems in which the relocation of a proton leads to zwitterionic structures. Similarly, the acetone exemplifies the case of those molecules whose tautomers undergo a reorganization of their electrons to maintain their neutrality. Contrary to conventional approaches, our CVs allow all the accessible tautomeric combinations and their reactive pathways to be explored in a single run without favouring any reactive candidates or having an initial guess on the reactivity of our system. Due to the plethora of chemical or biophysical processes involving tautomerisms, having structural and kinetics details of these processes at the atomistic level opens many possibilities for more specific and targeted approaches in several biochemical and medicinal chemistry applications. 

%Enzimatic activities, conformational problems and many other properties related to small peptides are strongly related to the presence of local charges along their structures. Identifying these thermodynamic states and assigning to them a statistical weight proportional to their stability is an useful tool not only for extracting acid-base equilibirum, but even for studying how they affect all their other properties.  
\begin{acknowledgement}
This research was supported by the European Union Grant No. ERC-2014-AdG-670227/VARMET. Calculations were carried out on the ETH Euler cluster.% and on the M\"{o}nch cluster at the Swiss National Supercomputing Center (CSCS).
\end{acknowledgement}
\section{Supporting Information}
\subsection{Collective Variables}
\begin{figure}
    \centering
    \includegraphics[width=.5\linewidth]{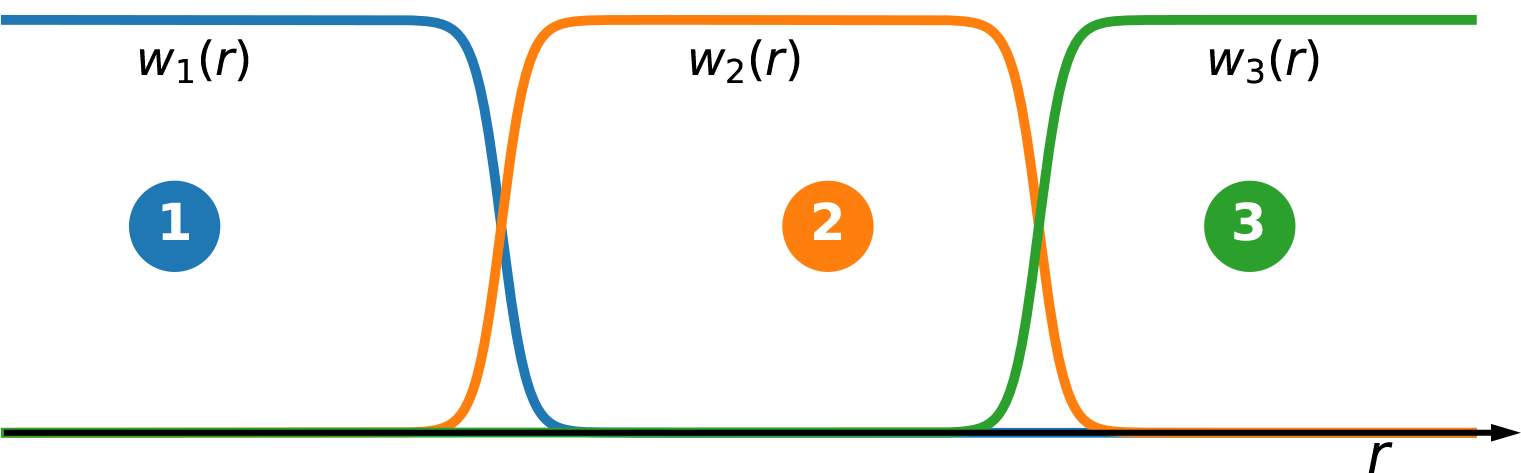}
    \caption{1D example of the Eq.~\ref{eq:omega}.}
    \label{fig:voronoi}
\end{figure}

According to the Voronoi principle, the total number of hydrogen atoms assigned to the site $i$ is taken as the fraction of protons that are much closer to this site than from all the other ones. Softmax functions allow to reproduce this behaviour without any discontinuity or singularity as shown in the Eq.~\ref{eq:omega} and Eq.~\ref{eq:rho},

\begin{equation}
    \omega_i(\mathbf{r})=\frac{e^{-\lambda |\mathbf{R}_i-\mathbf{r}|}}{\sum_m e^{-\lambda |\mathbf{R}_m-\mathbf{r}|}},
    \label{eq:omega}
\end{equation}

\begin{equation}
    \rho_i=\sum_{j \in H} \omega_i(\mathbf{R}_j),
    \label{eq:rho}
\end{equation}
where $\mathbf{R}$ is a vector of the atomic positions, the indexes $i$ and $m$ run over the atoms able to bond or release hydrogen atoms, $j$ runs over the hydrogen atoms and $\lambda$ is a parameter that controls steepness and selectivity of this function.
$\omega_i(r)$ approaches 1 when the atom $i$ is the closest to $r$ and otherwise its value approaches 0. (see Fig.~\ref{fig:voronoi}). In other words, it is a weight that says how much a particle in the position $r$ belongs to the Voronoi polyhedron of the site $i$.

Finally, the summation over the hydrogen atoms returns $\rho_i$, the total number of protons within the $i$th voronoi polyhedron.

\subsubsection{CV1: $s_p$}

\begin{figure}
    \centering
    \includegraphics[width=0.8\linewidth]{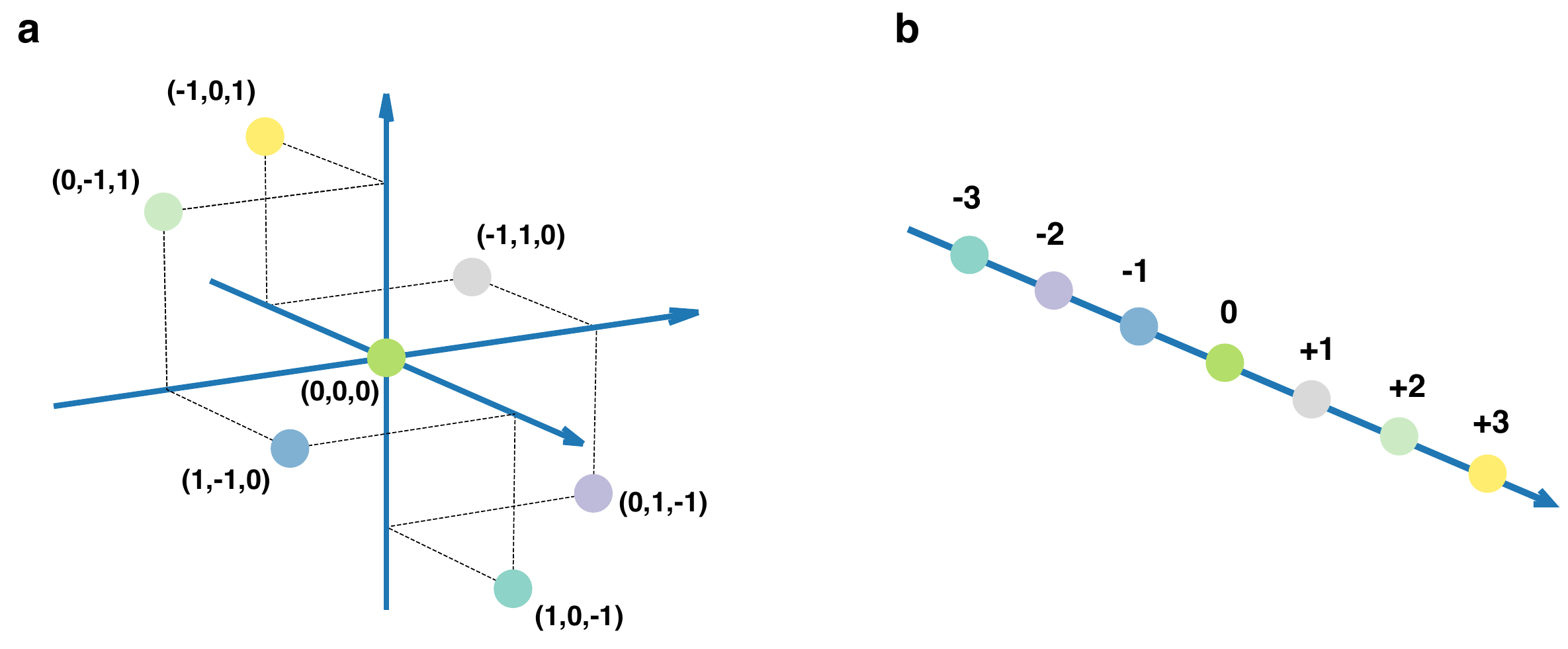}
    \caption{The 3D representation of vector $\Vec{q}$ (a) and its 1D projection as $s_p$ (b)}
    \label{fig:3d_projection}
\end{figure}

Once we have evaluated the instantaneous number of hydrogen atoms around each site, we compute their deviations from the respective reference (see Eq.~\ref{eq:delta}). Then, the overall excess or defect of protons for the solvent and every other moieties is taken as the summation of their site contributions (see Eq.~\ref{eq:q}). 

\begin{equation}
    \delta_i=\rho_i-\rho_i^0
    \label{eq:delta}
\end{equation}

\begin{equation}
    q_k=\sum_{i \in k}\delta_i
    \label{eq:q}
\end{equation}
The result of this operation is a vector $\Vec{q}=(q_0,q_1,\dots q_{N-1})$ with size equal to the number of inequivalent moieties and whose components indicate their overall excess or defect of protons. Finally, this vector is turned into a scalar through the dot product with another vector whose shape is $(2^0,2^1,2^2,\dots 2^{N-1})$. Since every component can assume only values between -1 and +1, this last transformation allows to assign any tautomeric combination to a different position in the CV space. Here is reported the example of a glycine molecule in aqueous solution, a system characterized by three different inequivalent moieties able to exchange protons and therefore showing seven theoretical protonation states (see Tab.~\ref{tab:s1}). Assuming we do not know anything about its reactivity, in principle every chemical group must be able to donate and accept a hydrogen atom. Since each thermodynamic state is described by a triplets of values, in this representation they can be represented as 3D vectors, see Fig.~\ref{fig:3d_projection}a and Tab.~\ref{tab:s1}.

The dot product with another vector $\Vec{X}$ collapses the multidimensional vector $\Vec{q}$ in a single scalar value. Another way of looking at this operation is like a linear combination

\begin{equation}
    s_p=\Vec{X}\cdot \Vec{q}=\sum_{k=0}^{N-1} X_k q_k,
\end{equation}
where $N$ is the vectors length and the coefficients $X_k$ can be chosen in order to provide a one-to-one correspondence between the scalar $s_p$ and the vector $\Vec{q}$. Since the components of $\Vec{q}$ can assume values in between -1 and +1, this is guaranteed when $\Vec{X}=(2^0,2^1,2^2)$, see Fig.~\ref{fig:3d_projection}b and Tab.~\ref{tab:s1}.

\begin{table}[htb]
    \centering
    \caption{The three components of the vector $\Vec{q}$ and the respective CV values.}
    $\begin{array}{rrr}
       q_0 & q_1 & q_2 \\
       \hline
        0 & 0 & 0  \\
        1 & -1 & 0  \\
        -1 & 1 & 0 \\
        0 & 1 & -1  \\
        0 & -1 & 1 \\
        1 & 0 & -1  \\
        -1 & 0 & 1  
    \end{array}
    \Longrightarrow
    \begin{tabular}{SSS}
        $X_{0} \cdot q_{0}$ &  $X_{1} \cdot q_{1}$ &  $X_{2} \cdot q_{2}$\\
       \hline
        0 & 0 & 0  \\
        1 & -2 & 0  \\
        -1 & 2 & 0 \\
        0 & 2 & -4  \\
        0 & -2 & 4 \\
        1 & 0 & -4  \\
        -1 & 0 & 4  
    \end{tabular}
    \Longrightarrow
    \begin{array}{r}
        s_p \\
       \hline
        0 \\
         -1 \\
         1 \\
         -2 \\
         2 \\
         -3 \\
         3 
    \end{array}$
    \label{tab:s1}
\end{table}

This ensures the possibility to explore all of its states starting from the most energetically accessible up to the highest one in energy. Moreover, this approach allows to address systems in which multiple and unknown competitive reactions are present without beforehand fix the reactive pairs.

\subsubsection{CV2: $s_d$}
\begin{figure}
    \centering
    \includegraphics[width=0.9\linewidth]{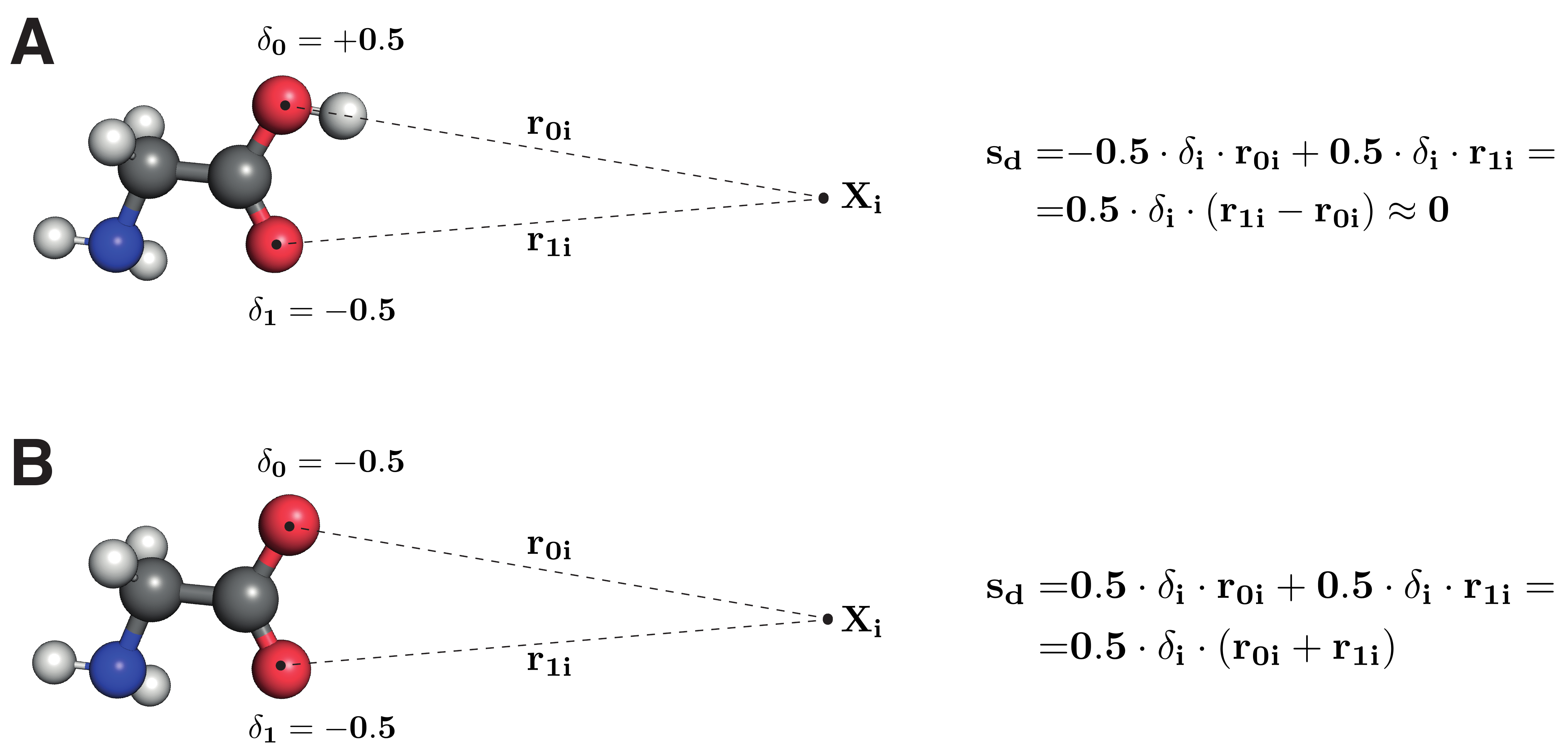}
    \caption{Schematic view of $s_d$ calculation between glycine (A) or glycinate (B), and a generic species $X_i$.}
    \label{fig:dist}
\end{figure}

With this CV we measure the distance between the sites that have exchanged a proton and, thus, not lying in their reference states. Here the reference value of protons in each Voronoi polyhedron is taken as the total number of protons assigned to the entire group $N_{H \in k}$ divided by the the total number of sites belonging to the $k$-th group, $N_{k}$. Starting from the Eq.~\ref{eq:rho}, the instantaneous deviation from its reference is computed as follow:

\begin{equation}
    \delta_i=\rho_{i \in k}-\frac{N_{H \in k}}{N_{k}}.
    \label{eq:delta_sd}
\end{equation}
In the example of the glycine molecule in aqueous solution, let us take the protolysis of its carboxylic group and consequent protonation of a water molecule. The water oxygen atoms have $N=32$ and $N_H=64$ while the two carboxylic oxygen atoms have $N=2$ and $N_H=1$. Before the reaction has taken place, every water oxygen atom has a value of $\rho$ approximately equal to 2, $\frac{N_H}{N}=2$ and therefore $\delta \approx 0$. After having subtracted a proton by the carboxylic group, one of the water oxygen site will have $\rho = 3$ and then $\delta = 1$. The same operation can be done for the two oxygen of the carboxylic group. Before the protolysis, assuming the hydrogen atom is bonded to the first carboxylic oxygen, the two sites have $\rho_0=1$ and $\rho_1=0$. Then their values of $\delta$ are +0.5 and -0.5 respectively. After the reaction has taken place both sites have $\delta$ equal to -0.5. This ensures that the opposite sign terms gives a zero contribution in the undissociated case (Fig.~\ref{fig:dist}A) and an averaged one in the other (Fig.~\ref{fig:dist}B).

\subsubsection{Restraint: $s_r$}
\begin{figure}
    \centering
    \includegraphics[width=0.8\linewidth]{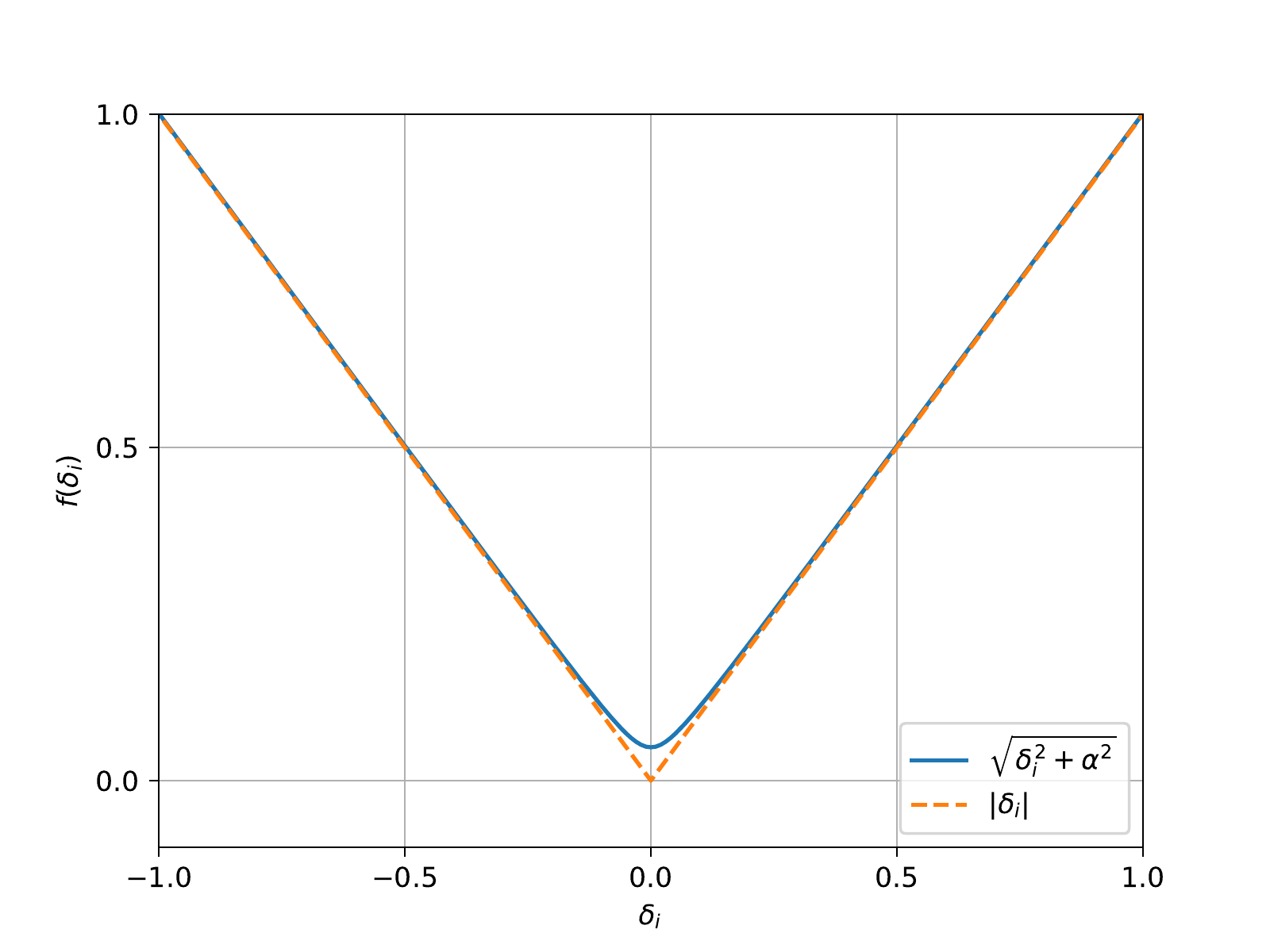}
    \caption{Different behaviour of absolute value function (orange dashed line) and the smoothed version (blue line) in proximity of $\delta_i = 0$. The parameter $\alpha$ controls the smoothness of the curve. In this plot the value of α has been set equal to $0.05$.}
    \label{fig:sr}
\end{figure}

In order to prevent simultaneous dissociation events, we restraint a third CV used just to monitor how many sites out of their reference states are present. The functional form of this CV is:

\begin{equation}
    s_r=\sum_i\sqrt{\delta_i^2+\alpha^2},
\end{equation}
where $i$ runs over the site indexes and $\alpha$ is a positive number much less than 1. With a proper value of $\alpha$ the square root term is a good approximation of the absolute value that allows to avoid the singularity for $\delta_i=0$ (see Fig.~\ref{fig:sr}). 
This CV returns the summation of the partial charge moduli and, by restraining it, we can limit at the given time the number of reacted pairs simultaneously present.

\subsubsection{Restraint2 (only acetone): distance}
In the acetone system, since the energy needed to break CH bonds is comparable to that for CC bonds, the external bias potential could in principle separate one of the two methyl groups from the main structure. In order to avoid this separation we applied a harmonic restraint to limit the maximum distance between bonded carbon atoms. The restraint potential $V_r$  prevents those atoms separation and its shape can be written as  

\begin{equation}
V_r(d)=
\begin{cases}
    k \cdot(d-d_0)^2 & d \ge d_0\\
    0 & d < d_0
\end{cases}
\end{equation}
where $k=50$, $d_0=1.65$ and $d$ are the carbon-carbon distances.

\subsection{FES along $\mathbf{s_p}$ and $\mathbf{s_d}$}

\begin{figure}
    \centering
    \includegraphics[width=.9\linewidth]{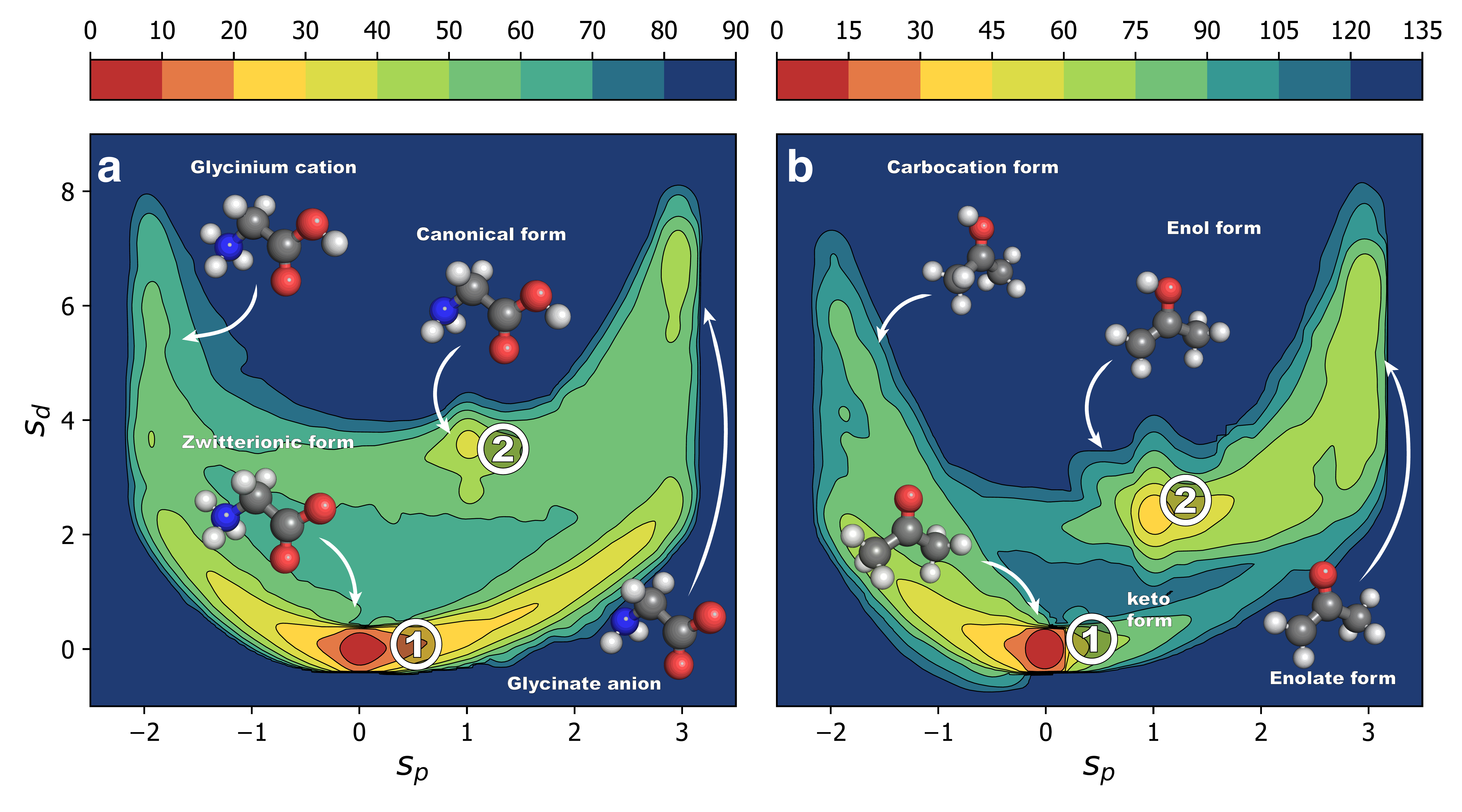}
    \caption{Free energy surfaces along $s_p$ and $s_d$ of glycine (panel a) and acetone (panel b) in aqueous solution. Color bars indicate the free energy expressed in \si{kJ \cdot mol^{-1}} units.}
    \label{fig:sp_sd}
\end{figure}

\subsection{Ab-initio MD setup}
Both the simulations have been set up as reported in Tab.~\ref{tab:md_setup}.

\begin{table}[htb]
    \centering
    \caption{Ab initio MD parameters.}
    \begin{tabular}{rrr}
    & \textbf{GLYCINE} & \textbf{ACETONE} \\
    \hline
    Reactive molecule & 1 & 1 \\
    Water molecules & 32 & 32 \\
    Ensemble & NVT & NVT  \\
    Temperaure (K) & 300 & 300 \\
    Thermostat & CSVR\cite{Bussi2007} & CSVR\cite{Bussi2007} \\
    Cell parameter (\AA) & 9.79 & 9.96 \\ 
    Basis sets & DZVP-MOLOPT-SR-GTH & DZVP-MOLOPT-SR-GTH \\
    Potential & GTH-PBE & GTH-PBE\\
    Energy cutoff (Ry) & 280 & 280\\
    Relative cutoff (Ry) & 40 & 40 \\
    EPS SCF & 1.0E-6 & 1.0E-6 \\
    XC Functional & PBE\cite{Perdew1996} & PBE\cite{Perdew1996}  \\
    Time step (fs) & 0.5 & 0.5 \\
    Length time (ps) & 375 & 450 \\
    \end{tabular}
    \label{tab:md_setup}
\end{table}

\subsection{Samples preparation}
Each system is composed by 32 water molecules and 1 of solute.
Both have been thermalized following these steps:
\begin{itemize}
    \item Geometry optimization,
    \item NVT MD simulation (1 ps),
    \item NPT MD simulation (10 ps),
    \item NVT MD simulation (2.5 ps);
\end{itemize}

\subsection{Well-Tempered Metadynamics setup}
Parameters adopted for PLUMED2 settings are reported in Tab.~\ref{tab:plumed_setup}.

\begin{table}[h]
    \centering
    \caption{PLUMED parameters.}
    \begin{tabular}{rrr}
    & \textbf{GLYCINE} & \textbf{ACETONE} \\
    \hline
    Gaussian hills heights & 1.0 & 2.0 \\
    Gaussian hills widths ($s_p$) & 0.075 & 0.075 \\ 
    Gaussian hills widths ($s_d$) & 0.4 & 0.4 \\ 
    Bias factor & 15 & 15 \\
    Temperature (K)& 300 & 300\\
    Hills deposition rate & 100 & 100 \\
    $\lambda$ ($s_p$) & 5 & 5 \\
    $\lambda$ ($s_d$) & 6 & 6 \\
    $\lambda$ ($s_r$) & 10 & 10\\
    $\alpha$ ($s_r$) & 1.0E-4 & 1.0E-4 \\
    \end{tabular}
    \label{tab:plumed_setup}
\end{table}

\bibliography{references}

\end{document}